\documentclass{aa}

\usepackage{graphicx}
\include{epsf}
\def\lesssim{\mathrel{\hbox{\rlap{\hbox{\lower4pt\hbox{$\sim$}}}\hbox{$<$}}}}
\def\gtrsim{\mathrel{\hbox{\rlap{\hbox{\lower4pt\hbox{$\sim$}}}\hbox{$>$}}}}

\begin{document}

%\thesaurus{07.19.1; 08.03.4; 08.16.5; 09.13.2}

%
        \title{On the master equation approach to diffusive grain-surface
        chemistry: the H, O, CO system}

     %   \subtitle{Deuterated Species and UV Shielding by Ambient Clouds}

        \author{T. Stantcheva,
               \inst{1}
	  V. I. Shematovich,
	        \inst{2}
               \and
               E. Herbst\inst{1,3}
               }

        \offprints{Eric Herbst, \email{herbst@mps.ohio-state.edu}}

        \institute{Department of Physics, The Ohio State University,
        Columbus OH 43210, USA
              \and
		  Institute of Astronomy of the Russian Academy of
Sciences, Moscow
		  109017, Russia
		  \and
                   Department of Astronomy, The Ohio State
                   University, Columbus, OH 43210, USA\\
                  }

        \date{Received; accepted}

\abstract{
We have used the master equation approach to study a moderately
complex network of diffusive reactions occurring on the surfaces of
interstellar dust particles.  This network is meant to apply to dense
clouds in which a large portion of the gas-phase carbon has already
been converted to carbon monoxide.  Hydrogen atoms, oxygen atoms, and
CO molecules are allowed to accrete onto dust particles and their
chemistry is followed.  The stable molecules produced are oxygen,
hydrogen, water, carbon dioxide (CO$_{2}$), formaldehyde (H$_{2}$CO),
and methanol (CH$_{3}$OH).  The surface abundances calculated via the
master equation approach are in good agreement with those obtained via
a Monte Carlo method but can differ considerably from those obtained
with standard rate equations.
}

\titlerunning{Master equation approach to grain chemistry}
\authorrunning{Stantcheva et al.}

\maketitle

\section{Introduction}

Rate equations have been widely used in simulations of gas-phase
processes in the interstellar medium (Le Teuff et al.~\cite{lmm2000}). 
Their application has also been extended to treat diffusive reactions
on the surfaces of dust particles (Pickles \& Williams \cite{pw1977};
Hasegawa et al.~\cite{hhl1992}).  This approach, however, is valid
only if the average number of reactive surface species per grain is
large and the discrete nature of the system can be neglected.  When
this number becomes small, the rate equations may no longer constitute
an accurate description of the chemical processes (Tielens \& Hagen
\cite{th1982}); moreover, for species with surface populations less
than one, it can become meaningless to use the rate equations
(Charnley et al.~\cite{ctr1997}; Caselli et al.~\cite{chh1998}).

This problem has spurred attempts to develop alternative methods for
handling diffusive reactions on granular surfaces.  One possible
solution is to use Monte Carlo procedures to simulate the
grain-surface chemistry.  Different methods, based on Monte Carlo
simulations, have already been employed for various grain surface
networks (Tielens \& Hagen \cite{th1982}; Charnley et
al.~\cite{ctr1997}; Charnley \cite{c2001}; Caselli et
al.~\cite{css2002}).  These simulations were performed under the
constraint that, during the evolution of the surface chemistry, the
gas-phase abundances of all species stay constant - a requirement that
cannot be met in complex gas-grain models, where it is essential that
the gas-phase and the grain-surface reactions run in a parallel mode
during the calculations and that both adsorption onto and desorption
from grain surfaces occur (Ruffle \& Herbst \cite{rh2000}).
It does not appear possible, however, to follow simultaneously the
gas-phase chemistry with rate equations and the surface chemistry with
a Monte Carlo approach.  Although one can
     use the Monte Carlo method to simulate gas-phase
reactions (Charnley \cite{ch1998}), a Monte Carlo solution
for both gas-phase and grain-surface chemisty could be done only with
an exceedingly large amount of computing time, and has not
yet been attempted.  Current gas-grain chemical
models (e.g.  Ruffle \& Herbst \cite{rh2000}) use a semi-empirical
modification of the standard rate equations for diffusive surface chemistry
known as the "modified rate approach" (Caselli et al.~
\cite{chh1998}; Stantcheva et al.~\cite{sch2001}; Caselli
et al.~\cite{css2002}).  Although this approach is efficient, its
semi-empirical nature raises doubts of its suitability under all
conditions.

Recently, two groups (Biham et al.~\cite{bfp2001}; Green et
al.~\cite{gtp2001}) proposed a second stochastic approach to granular
chemistry, known as the master equation treatment.  In this approach,
differential rate equations for species with a small surface abundance
are replaced by differential equations in which one solves for the
probability that a specific number of atoms or molecules of that
species (0,1,\ldots) is present on a grain at any time.  In general,
the probabilities for each surface species are not independent and so an exact
treatment requires the determination of joint probabilities (e.g.,  for
0 of species A, 1 of species B, etc.).  Green et al.~(\cite{gtp2001})
used the master equation method to study the simple O, H system; in
this system O and H atoms, with fixed gas-phase abundances, are
allowed to adsorb onto grains and react to produce the three molecules
OH, O$_{2}$, and H$_{2}$ via diffusive processes.  For this system,
determination of the surface populations of the minor but important
species H and O requires the calculation of two-body probabilities,
while surface abundances of the diatomic species can be calculated
from normal rate equations.  The two-body probability approach to the
simple O, H system can itself be simplified by a number of approximate
methods (Green et al.  \cite{gtp2001}).  One approximation, proposed
by Biham et al.~(\cite{bfp2001}) but not attempted by Green et
al.~(\cite{gtp2001}), reduces the two-body probability to independent
probabilities for the individual species O and H. With this approach, the
so-called ``many-body'' master equation reduces to separate
master equations for the individual species.  Although the
approximation leads to a significantly smaller number of simultaneous
differential equations to solve, its validity is not clear for systems
with strong correlations between the surface abundances of the
different minor species.

If one attempts to scale up the many-body master equation approach to
model realistic complex networks of surface reactions, one can
experience serious problems involving both computer time and memory,
unless suitable approximations can be found.  In order to learn more
about the use of the method for larger systems, we have considered an
intermediate system more complex than previously studied but not at
the level of complexity needed for a complete network of surface
reactions.  In particular, we report here the use of the many-body
master equation approach to solve a grain-surface chemical network in
which gaseous H, O, and CO accrete onto grain surfaces and lead to the
production of molecular hydrogen, formaldehyde, methanol, water,
oxygen, and carbon dioxide (Caselli et al.~\cite{css2002}; see also
Charnley et al.~\cite{ctr1997}).  This system has previously been
studied by the modified rate and Monte Carlo approaches (Caselli et
al.~\cite{css2002}).  We consider here a variety of different
diffusive rates and temperatures to see how the master equation
approach fares.  As opposed to the previous treatments based on the
master equation, we utilize a
time-dependent approach designed eventually to be coupled with a
time-dependent gas-phase chemistry, since the advantage of the direct
solution of the master equation compared with the Monte Carlo
realization is that the differential equations are easily coupled to
rate equations for the gas-phase species.

The organization of the paper is as follows.  In the next section, we
discuss the chemical network of diffusive surface reactions and the
different rates chosen.  In Sect.~3, we write out the differential
equations needed to model the methanol system via the master equation
approach, while in Sect.~4 we present our results, and compare them
with Monte Carlo and rate approaches.  A general discussion is
contained in Sect.~5.  In addition, a detailed mathematical
discussion of the master equation and Monte Carlo realizations to
stochastic kinetics is contained in the Appendix.
%__________________________________________________________________

\section{The H, O, CO Network}

In this network, three gas-phase species -- H, O, and CO -- are allowed
to accrete onto a granular surface where they react via diffusion to produce
the stable molecules H$_{2}$, O$_{2}$,
H$_{2}$O, H$_{2}$CO, CH$_{3}$OH, and CO$_{2}$, as well as reactive intermediate
species.  Of the ten reactions, nine are association reactions, in
which a single product is formed.  No gas-phase
chemistry is allowed to occur and the gas-phase concentrations of the
three species are assumed to remain constant despite accretion onto
grains.  This somewhat contradictory assumption permits the surface
chemistry to occur only over a period sufficiently short that the
abundances of the gas-phase species do not change dramatically.

The calculations were carried out for three different sets of
gas-phase abundances of the accreting species, known respectively as
the low, intermediate, and high density cases.  All of these cases
refer to dense clouds in which most of the atomic hydrogen has already
been converted into molecular hydrogen, leaving only a small remnant
in atomic form.  Such conditions pertain when a significant amount of
CO has been produced via gas-phase chemistry.  The abundances
$n$(cm$^{-3}$) of H, O, and CO shown in Table~\ref{density} were
obtained from steady-state gas-phase models run at molecular hydrogen
densities of 10$^{3}$, 10$^{4}$, and 10$^{5}$ cm$^{-3}$.  In the
low-density case, there is more atomic hydrogen than O and CO, and the
chemistry will be seen to be strongly reductive, whereas in the
high-density case, there is little atomic hydrogen around to react
with CO.

\begin{table}
\caption{H, O, and CO gas-phase abundances (cm$^{-3}$) utilized}
\label{density}
\begin{flushleft}
\begin{tabular}{clll}
\hline \hline
     \noalign{\smallskip}

Abundance $n$ & Low & Intermediate & High \\
\hline
	H & 1.15 & 1.15 & 1.10 \\
	O & 0.09 & 0.75 & 7.0\\
	CO & 0.075 & 0.75 & 7.5\\
                \hline \hline
			 \noalign{\smallskip}
        \end{tabular}
        \end{flushleft}
       \end{table}

     The surface reactions for the network are listed in Table~\ref{methr}
     along with activation energies in K where appropriate.  The
     activation energies $E_{\rm a}$ (K) are approximate only (Caselli et
     al.~\cite{css2002}).  The total number of surface species is 12; this
     includes the highly reactive radicals OH, HCO, and H$_{3}$CO. The key
     reaction sequence in the network is the relatively slow hydrogenation
     of CO into methanol (CH$_{3}$OH) via H-atom addition reactions.
     Unambiguous laboratory evidence for this hydrogenation is not
     available (Hiraoka et al.~\cite{hst2000}),
     but conditions in interstellar clouds are not those in the laboratory
     and the complete synthesis in Table~\ref{methr} is by no means ruled
     out by experiments.   Once produced, all stable species
     except H$_{2}$ remain on the grain
surface; the evaporation of molecular hydrogen is included.

       \begin{table}
	\caption[]{Surface reactions in the H,O,CO model}
	\label{methr}
	\begin{flushleft}
	\begin{tabular}{lll}
                \hline \hline
                \noalign{\smallskip}
	Number & Reaction & $E_{\rm a}$ (K)    \\
                \noalign{\smallskip}
                \hline
	1  & H + H$\longrightarrow$ H$_{2}$             &      \\
	2  & H + O$\longrightarrow$ OH                  &      \\
	3  & H + OH$\longrightarrow$ H$_{2}$O           &      \\
	4  & H + CO$\longrightarrow$ HCO                & 2000 \\
	5  & H + HCO$\longrightarrow$ H$_{2}$CO         &      \\
	6  & H + H$_{2}$CO$\longrightarrow$ H$_{3}$CO   & 2000 \\
	7  & H + H$_{3}$CO $\longrightarrow$ CH$_{3}$OH &      \\
	8  & O + O $\longrightarrow$ O$_{2}$            &      \\
	9  & O + CO $\longrightarrow$ CO$_{2}$          & 1000 \\
	10 & O + HCO $\longrightarrow$ CO$_{2}$+H       &      \\
                \noalign{\smallskip}
                \noalign{\smallskip}
                \hline \hline
             \end{tabular}
		 \end{flushleft}
       \end{table}

Whether one uses rate equations, the Monte Carlo approach, or the
direct master equation method, it is necessary to utilize diffusion
rate coefficients $k$ for the reactive surface species (Hasegawa et al.
~\cite{hhl1992}; Appendix).  The rate coefficients here are in
units of s$^{-1}$, as preferred by Caselli, Hasegawa, \& Herbst
(\cite{chh1998}).  These are  the sum of the rates
$t^{-1}_{\rm diff}$ (s$^{-1}$) of the reactive partners to traverse an
entire grain, which is here assumed to contain 10$^{6}$ binding sites,
multiplied by a factor $\kappa$ that accounts for any non-zero
chemical activation barrier (Hasegawa et al.~\cite{hhl1992}).  The
rates depend strongly on the barriers against diffusion $E_{\rm b}$
from site to site chosen, and whether diffusion occurs via thermal
hopping or via quantum mechanical tunneling (Tielens \& Hagen
\cite{th1982}).

In our calculation, we have considered three sets of barriers against
diffusion.  The first, which comes from the earlier astrochemical
literature (Allen \& Robinson \cite{ar1977}; Tielens \& Hagen
\cite{th1982}; Hasegawa et al.~\cite{hhl1992}), contains rather low
barriers and allows efficient tunneling for atomic H.
The second and third are based on the recent experiments of Pirronello
et al.~ (\cite{pbl1997}, \cite{plr1999}) as simulated by Katz et al.~
(\cite{kfb1999}), which show that atomic H moves much more slowly on
olivine and amorphous carbon than previously assumed in the
astrochemical literature.  Two sets of barriers based on these
experiments on olivine have been used (Ruffle \& Herbst \cite{rh2000})
- one, designated ``slow H'', in which only the H atom barrier is
raised, and the other, designated ``slow'', in which all other barriers
are raised proportionately.  In both sets of barriers, no tunneling is
allowed, since no tunneling of H was detected in the laboratory.  For
our calculations here, the earlier astrochemical values are used
principally because the slower diffusion rates cannot produce much
formaldehyde and methanol in the small times considered, and so are
not emphasized (Caselli et al.~\cite{css2002}).  In our more complex
gas-grain models, the slower rates have been used, and show that
formaldehyde and methanol can be produced over long periods of time
(Ruffle \& Herbst \cite{rh2000}).

In addition to diffusive rates, the rates of adsorption and desorption
must be included in our calculation (Hasegawa et al.  \cite{hhl1992}).
Adsorption is assumed to occur at unit efficiency once a gas-phase
species strikes a grain.  We consider only
thermal desorption (evaporation) and treat it as in previous models
(see e.g. Caselli et al.~\cite{chh1998}); the rates, exponentially
dependent on the desorption energy $E_{\rm D}$, are included for the
accreting gas-phase species and for molecular hydrogen product.  Heavy
molecular species desorb too slowly for this process to be considered
here.  The small barriers against diffusion and the desorption
energies for all species in the model are listed in
Table~\ref{barriers}.

Table~\ref{rates} gives our values for the accretion rate coefficients
$k_{\rm acc}$
(cm$^{3}$ s$^{-1}$) onto a grain, the evaporation rates $t^{-1}_{\rm
evap}$ from the grain, and the diffusion rates for the species H, O,
and CO at 10 K, unless they are vanishingly small.
Calculations have been done mainly at this temperature,
although temperatures up to 20 K have been considered.

       \begin{table}
          \caption{Energy barriers against diffusion (low values) and
          desorption energies (K)}
             \label{barriers}
         \begin{tabular}{lcc}
                \hline \hline
                \noalign{\smallskip}
                Species  &  $E_{\rm b}$ (K) & $E_{\rm D}$ (K) \\
                \noalign{\smallskip}
                \hline
                \noalign{\smallskip}
H           & 100 & 350  \\
O          & 240 & 800  \\
OH         & 378 & 1260 \\
H$_{2}$    & 135 & 450  \\
O$_{2}$    & 363 & 1210  \\
H$_{2}$O   & 558 & 1860  \\
CO         & 363 & 1210  \\
HCO        & 453 & 1510  \\
H$_{2}$CO  & 528 & 1760  \\
CH$_{3}$O  & 651 & 2170 \\
CH$_{3}$OH & 618 & 2060 \\
CO$_{2}$   & 750  & 2500 \\
                \noalign{\smallskip}
                \hline \hline
             \end{tabular}
%\begin{list}{}{}
%\item[$^{\mathrm{a}}$] Both barriers from Allen \& Robinson
%(\cite{ar1977}); others from Hasegawa et al. (\cite{hll1992}) CHECK -
%may not be so clear cut.
%\end{list}
       \end{table}

       \begin{table}
          \caption{Assorted rates for selected species at 10 K}
             \label{rates}
             \begin{tabular}{llll}
                \hline \hline
                \noalign{\smallskip}
                Species  & $k_{\rm acc}$ (cm$^{3}$s$^{-1}$)
		& $t_{\rm evap}^{-1}$ (s$^{-1}$)
		& $t_{\rm diff}^{-1}$ (s$^{-1}$) \\
                \noalign{\smallskip}
                \hline
                \noalign{\smallskip}
H          & 1.45(-5)& 1.88(-3) & 5.14(+4)$^{\mathrm{a}}$ \\
O          & 3.62(-6)& 2.03(-23) & 4.24(-5)\\
CO         & 2.73(-6) & &\\
                \noalign{\smallskip}
                \hline \hline
             \end{tabular}
\begin{list}{}{}
\item[$^{\mathrm{a}}$] Quantum tunelling included
\end{list}
       \end{table}

%************************

\section{Master equation for H,O,CO system}

Of the three species - H, O, and CO - that accrete onto grain
surfaces, the first two are very reactive and never build up large
surface populations.  On the other hand, CO reacts only slowly via
reactions with activation energy and so can build up a large surface
population under certain circumstances.  In general, all surface
species in our network can be classified as either major or minor
species.  The major ones correspond to the stable species that react
slowly if at all (CO, H$_{2}$CO, CH$_{3}$OH, H$_{2}$O, O$_{2}$,
H$_{2}$, CO$_{2}$) and can build up large abundances, while the minor
ones are atoms and radicals likely to have a surface number, defined
as the number of species per grain, at or below unity.  We treat the
minor species - H, O, OH, HCO, and H$_{3}$CO - probabilistically with
corresponding surface numbers {\it i$_{1}$}, {\it i$_{2}$}, {\it
i$_{3}$}, {\it i$_{4}$}, and {\it i$_{5}$}.

The first step in the master equation approach is to solve for the
joint probability {\it P}({\it i$_{1}$},{\it i$_{2}$},{\it
i$_{3}$},{\it i$_{4}$},{\it i$_{5}$}), defined as the probability that
{\it i$_{1}$, \it i$_{2}$}, {\it i$_{3}$}, {\it i$_{4}$}, and {\it
i$_{5}$} numbers of minor species exist on the surface as a function of
time.  In the calculations discussed here, we start with the initial
condition that the joint probability is unity for $P(0,0,0,0,0)$.  Let
X and Y be, respectively, the j$^{th}$ and k$^{th}$ reactive elements of the
ordered set \mbox{\{H, O, OH, HCO, H$_{3}$CO \}}.  Let Z represent
any of the major species. The time derivative
of the five-body probability for each value of
$i_{1},i_{2},i_{3},i_{4},i_{5}$ can then be written as (Appendix)

%\begin{eqnarray}
\begin{equation}
\label{prob}
\begin{array}{l}
\frac{{\textstyle dP}}{{\textstyle dt}}(i_{1},i_{2},i_{3},i_{4},i_{5}) = \\ \\
	\displaystyle{\sum_{\{\mathrm {X}\}}} k_{\rm acc}(\mathrm{X})n(\mathrm{X})
	\left[ P(...,i_{j}-1,...) - P(...,i_{j},...)\right] \\ \\
	+ \displaystyle{\sum_{\{\mathrm{X}\}}} t_{\rm evap}^{-1}(\mathrm{X})
	\left[(i_{j}+1)P(...,i_{j}+1,...) -
	i_{j}P(...,i_{j},...)\right] \\ \\
	+ \displaystyle{\sum_{\{\mathrm{X,Y}\}}} k_{\mathrm{X},\mathrm{Y}}
	(i_{j}+1)(i_{k}+1)P(...,i_{j}+1,...,i_{k}+1,...) \\ \\
	- \displaystyle{\sum_{\{\mathrm{X,Y}\}}} k_{\mathrm{X},\mathrm{Y}}
	(i_{j})(i_{k})P(...,i_{j},...,i_{k},...) \\ \\
	+ \displaystyle{\sum_{\{\mathrm{X}\}}} k_{\mathrm{X},\mathrm{X}}
	\frac{(i_{j}+2)(i_{j}+1)}{2}P(...,i_{j}+2,...) \\ \\
	- \displaystyle{\sum_{\{\mathrm{X}\}}} k_{\mathrm{X},\mathrm{X}}
	\frac{i_{j}(i_{j}-1)}{2}P(...,i_{j},...) \\ \\
	+\displaystyle {\sum_{\{\mathrm{X,Z}\}}}\langle N_{\mathrm{Z}} \rangle
	 k_{\mathrm{X}, \mathrm{Z}}
	\left[(i_{j}+1)P(...,i_{j}+1,...) \right] \\ \\
	 - 	\displaystyle {\sum_{\{\mathrm{X,Z}\}}}\langle
N_{\mathrm{Z}} \rangle
	 k_{\mathrm{X}, \mathrm{Z}} \left[i_{j}P(...,i_{j},...)\right] .
%\end{eqnarray}
\end{array}
\end{equation}
\noindent
where $n$ stands for gas-phase concentration and $\langle N \rangle$ for
surface abundance; i.e., the average number of atoms or molecules per grain of
a species.

The first term on the right-hand side of eq.~(\ref{prob}) accounts for
changes in the state of the surface on a particular grain due to
accretion of species.  In this particular case, the sum consists of
two terms because the only accreting species with minor surface
abundance are H and O. The second term describes the changes of the
system due to evaporation, and the remaining terms take into account
any changes due to surface reactions.  These terms are subdivided into
expressions for reaction between two distinct minor species (X,Y), for
self-reaction (X,X), and for reaction between a minor and a major
species (X,Z).  Note that all of the terms refer to reactants;
there is also one minor product - H atoms in reaction 10 of Table
\ref{methr}.  To include the production of H in Eq.~(\ref{prob})
requires a term which, when X and Y are equal to O and HCO,
contains probability functions where three indices change.  We have
not included this term in eq.~ (\ref{prob}) for simplicity, but it is
of course included in our calculations.  The average abundances of the
major species are obtained from rate equations discussed below.  It is
easily shown that the total probability as a function of time is
conserved at unity.

Because the abundances of minor species
are low, joint probabilities with high numbers of these particles are
very unlikely and therefore, the probabilities for such states can be
neglected.  In particular, we choose a set of parameters $\mathcal{N}$
$=$
\{$\mathcal{N}_{1},\mathcal{N}_{2},\mathcal{N}_{3},\mathcal{N}_{4},\mathcal{N}_{5}$
\} such that the only probabilities {\it P}(...,$i_{j}$,...)  to be
considered possess $i_{j}\leq \mathcal{N}_{j}$,
\textit{j}$\in\{1,2,3,4,5\}$.  Specific choices for the set
$\mathcal{N}$ are discussed in the Results section.  It is obvious that
for the master equation method to be feasible, the set $\mathcal{N}$
must contain elements as small as possible.
%for each \textit{j} all probabilities {\it
%P}(...,$i_{j}$,...) were neglected when $i_{j}>\mathcal{N}_{j}$.

Once the probabilities are determined by integration for a specific
time, the average numbers of minor species, $\langle
N_{\mathrm{H}}\rangle$, $\langle N_{\mathrm{O}}\rangle$, etc., as well
as the correlation terms $\langle
N_{\mathrm{H}}N_{\mathrm{O}}\rangle$, $\langle
N_{\mathrm{H}}N_{\mathrm{OH}}\rangle$, etc., can be calculated from
these probabilities; e.g.,
\begin{equation}
\langle N_{\mathrm{H}}\rangle = {\displaystyle \sum^{\mathcal
		{N}}_{\scriptstyle{i_{1}, i_{2},i_{3}, i_{4}, i_{5}}}}
	i_{1} P(i_{1},i_{2},i_{3},i_{4},i_{5}).
\end{equation}
Both average numbers and correlations are then used in the rate equations for
the abundance of the major surface species, while the latter can also be
used to test how independent or correlated the minor species are.
In the rate equations for major species,
shown below, the division of species into major and minor ones leads
to the fact that correlations are only used for pairs of minor species:

\begin{equation}
\label{reH2}
\begin{array}{l}
\frac{{\textstyle d \langle
	N_{\mathrm{H}_{2}}\rangle}}{{\textstyle dt}} =
	- \; t_{\rm evap}^{-1}(\mathrm{H}_{2}) \; \langle N_{\mathrm{H}_{2}}
	\rangle \\ \\
	\; \; \; \; \;
	+ \; k_{\mathrm{H},\mathrm{H}} \times 0.5 \times \langle
	N_{\mathrm{H}}(N_{\mathrm{H}}-1)\rangle,
\end{array}
\end{equation}

\begin{equation}
\label{reO2}
\frac{{\textstyle d \langle N_{\mathrm{O}_{2}}\rangle}}{{\textstyle dt}} =
	k_{\mathrm{O},\mathrm{O}} \times 0.5 \times \langle
	N_{\mathrm{O}}(N_{\mathrm{O}}-1)\rangle,
\end{equation}

\begin{equation}
\label{reH2O}
\frac{{\textstyle d \langle
	N_{\mathrm{H}_{2}\mathrm{O}}\rangle}}{{\textstyle dt}} =
	k_{\mathrm{H},\mathrm{OH}} \times  \langle
	N_{\mathrm{H}}N_{\mathrm{OH}}\rangle,
\end{equation}

\begin{equation}
\label{reCO}
\begin{array}{l}
\frac{{\textstyle d \langle N_{\mathrm{CO}}\rangle}}{{\textstyle dt}}=
	k_{\rm acc}(\mathrm{CO}) n(\mathrm{CO}) \\ \\
	\; \; \; \; \;
	- \; k_{\mathrm{H},\mathrm{CO}} \langle N_{\mathrm{CO}} \rangle
	\langle N_{\mathrm{H}} \rangle
	- \; k_{\mathrm{O},\mathrm{CO}} \langle N_{\mathrm{CO}} \rangle
	  \langle N_{\mathrm{O}} \rangle
\end{array}
\end{equation}

\begin{equation}
\label{reH2CO}
\begin{array}{l}
\frac{{\textstyle d \langle
N_{\mathrm{H}_{2}\mathrm{CO}}\rangle}}{{\textstyle dt}} =
	k_{\mathrm{H},\mathrm{HCO}}\langle
N_{\mathrm{H}}N_{\mathrm{HCO}} \rangle  
	\\ \\
	\; \; \; \; \;
	- \; k_{\mathrm{H},\mathrm{H}_{2}\mathrm{CO}}
	\langle N_{\mathrm{H}_{2}\mathrm{CO}} \rangle
	 \langle N_{\mathrm{H}} \rangle
\end{array}
\end{equation}

\begin{equation}
\label{reCH3OH}
\frac{{\textstyle d \langle
	N_{\mathrm{CH}_{3}\mathrm{OH}}\rangle}}{{\textstyle dt}} =
	k_{\mathrm{H},\mathrm{H}_{3}\mathrm{CO}}
	\langle N_{\mathrm{H}}N_{\mathrm{H_{3}CO}} \rangle,
\end{equation}

\begin{equation}
\label{reCO2}
\begin{array}{l}
\frac{{\textstyle d \langle N_{\mathrm{CO}_{2}}\rangle}}{{\textstyle dt}} =
	k_{\mathrm{O},\mathrm{CO}}
	\langle N_{\mathrm{CO}} \rangle\langle N_{\mathrm{O}} \rangle
	+ k_{\mathrm{O},\mathrm{HCO}}
	\langle N_{\mathrm{O}}N_{\mathrm{HCO}} \rangle.
\end{array}
\end{equation}

-----------------------

\subsection{Approximation of Independent Probabilities}

Considering the large number of coupled differential equations for the
many-body $P$ required in the exact master equation approach, it is
perhaps useful to consider whether the use of one-particle
probability functions is adequate.  This approach was suggested but not
tested by
     Biham et al. ~(\cite{bfp2001}) for the simple O, H system, and leads to
     a different and
somewhat simpler system of differential equations. The H, O, CO
system reduces to the O, H system if no CO is allowed to accrete onto
grains and the OH radical is treated as unreactive.  Biham et
al.~(\cite{bfp2001}) used the notation
{\it P}$_{\mathrm H}$({\it i}) and {\it P}$_{\mathrm O}$({\it j}) for
the (independent) probabilities that {\it i} H atoms and {\it j} O atoms
are on the
surface, respectively. For X,Y $\in$ $\{$H, O$\}$, the equation for
the probability that species X has
{\it i} atoms  is

\begin{equation}
\begin{array}{l}
\frac{{\textstyle dP}_{\mathrm{X}}}{{\textstyle dt}}(i) =
k_{\rm acc}(\mathrm{X})n(\mathrm{X})[P_{\mathrm{X}}(i-1) -
P_{\mathrm{X}}(i)] \\ \\
+ t_{\rm evap}^{-1}(\mathrm{X})[ (i+1)P_{\mathrm{X}}(i+1) -
iP_{\mathrm{X}}(i)] \\ \\
+ k_{\mathrm{X},\mathrm{X}}
\left[\frac{(i+2)(i+1)}{2}P_{\mathrm{X}}(i+2) -
\frac{i(i-1)}{2}P_{\mathrm{X}}(i)\right] \\ \\
%+ \displaystyle{\sum_{\{\mathrm{Y}\}}}k_{\mathrm{X},\mathrm{Y}}
+ k_{\mathrm{X},\mathrm{Y}}\langle N_{\mathrm{Y}}\rangle
[(i+1)P_{\mathrm{X}}(i+1) - iP_{\mathrm{X}}(i)] \;
\end{array}
\end{equation}

\noindent with initial conditions: {\it P}$_{\mathrm H}$(0)=1 and
{\it P}$_{\mathrm O}$(0)=1. Note that species X depends on species Y
only through its average abundance.

The rate equations for the major species
are in the form

\begin{eqnarray}
\frac{d \langle N_{\mathrm{OH}}\rangle}{dt} & =
	& k_{\mathrm{O},\mathrm{H}} \times
	\sum^{\infty}_{i_{1}=1}i_{1}P_{\mathrm{H}}(i_{1}) \times
	\sum^{\infty}_{i_{2}=1}i_{2}P_{\mathrm{O}}(i_{2}) \\
	\nonumber
	& = & k_{\mathrm{O},\mathrm{H}} \times \langle N_{\mathrm{H}}
	\rangle \langle N_{\mathrm{O}}  \rangle \; , \\ 
%	\nonumber \\
\frac{d \langle N_{\mathrm{H}_{2}}\rangle}{dt} & =
	& k_{\mathrm{H},\mathrm{H}} \times
	\sum^{\infty}_{i_{1}=2}\frac{i_{1}(i_{1}-1)}{2}P_{\mathrm{H}}(i_{1})
	\\ \nonumber
	& = & k_{\mathrm{H},\mathrm{H}} \times \frac{ \langle
	N_{\mathrm{H}}(N_{\mathrm{H}}-1)\rangle}{2} \; , \\ 
%	\nonumber \\
\frac{d \langle N_{\mathrm{O}_{2}}\rangle}{dt} & =
	& k_{\mathrm{O},\mathrm{O}} \times
	\sum^{\infty}_{i_{2}=2}\frac{i_{2}(i_{2}-1)}{2}P_{\mathrm{O}}(i_{2})
	\\ \nonumber
	& = & k_{\mathrm{O},\mathrm{O}} \times \frac{\langle 
	N_{\mathrm{O}}(N_{\mathrm{O}}-1)\rangle}{2} \; .
\end{eqnarray}
%where the right-hand sides of these equations can also
%be written in terms of average surface abundances $\langle
%N_{\ldots}\rangle$.
For the production of OH, the average abundances of O and H
appear as independent products rather than as a correlation, which
would be the case if the minor (H, O) species were determined with a
two-body probability function.

%************************

%************************
\subsection{Implementation}

The calculations were performed with the use of a Gear algorithm on a
Cray SV1 computer.  To enhance the performance, the equations were
supplied by a subroutine which wrote them in an explicit form.
Calculations were virtually all performed to a time of 1000 yr, which
is rather short by astronomical scales, but is more than sufficient to
allow the minor species to reach a steady-state condition and the
major species to increase linearly with time.  We have investigated
how the needed computer time depends on the astronomical time scale
and find that for the system studied, there is hardly any difference
in computer time if the astronomical time is increased 100-fold.  To
increase the stability of the calculation, after every call of the
equation-solving routine, the total sum of the many-body probabilities
was evaluated and the amount by which it deviated from unity was added
to the probability for the state with no minor species.

\section{Results}

Unless we state that slow H or slow rates are being used, the results
below are for the fast rates discussed in Sect.~2.  The fast
diffusion rates are emphasized both because they allow the rapid
production of methanol and because they cause a larger discrepancy
between the results of the standard rate equations and more exact
methods.  Unless stated to the contrary, the temperature is fixed at
10 K.

%************************
\subsection{Check of the Independence Approximation}

Before proceeding to our main results, it is interesting to check the
validity of the suggestion by Biham et al.~(\cite{bfp2001}) that
independent probabilities be utilized.  We have used the simple O,H
system for a comparison among the following five methods: the Monte
Carlo approach (Charnley \cite{c2001}; Appendix), the master equation approach,
the approximation to
the master equation approach of Biham et al.~(\cite{bfp2001}), the
rate equation approach, and the modified rate equation approach
(Stantcheva et al.~\cite{sch2001}). Calculations were performed at a
fixed concentration of gas-phase atomic oxygen (1 cm$^{-3}$) and a variable
concentration of gas-phase atomic hydrogen.  No desorption of the products was
allowed.  In the master equation calculations, the
maximum allowed numbers $\{\mathcal{N}_{i}\}$ for H and O
were never larger than 5.

\begin{figure}
       \centering
       \rotatebox{-90}{\includegraphics[width=20 em]{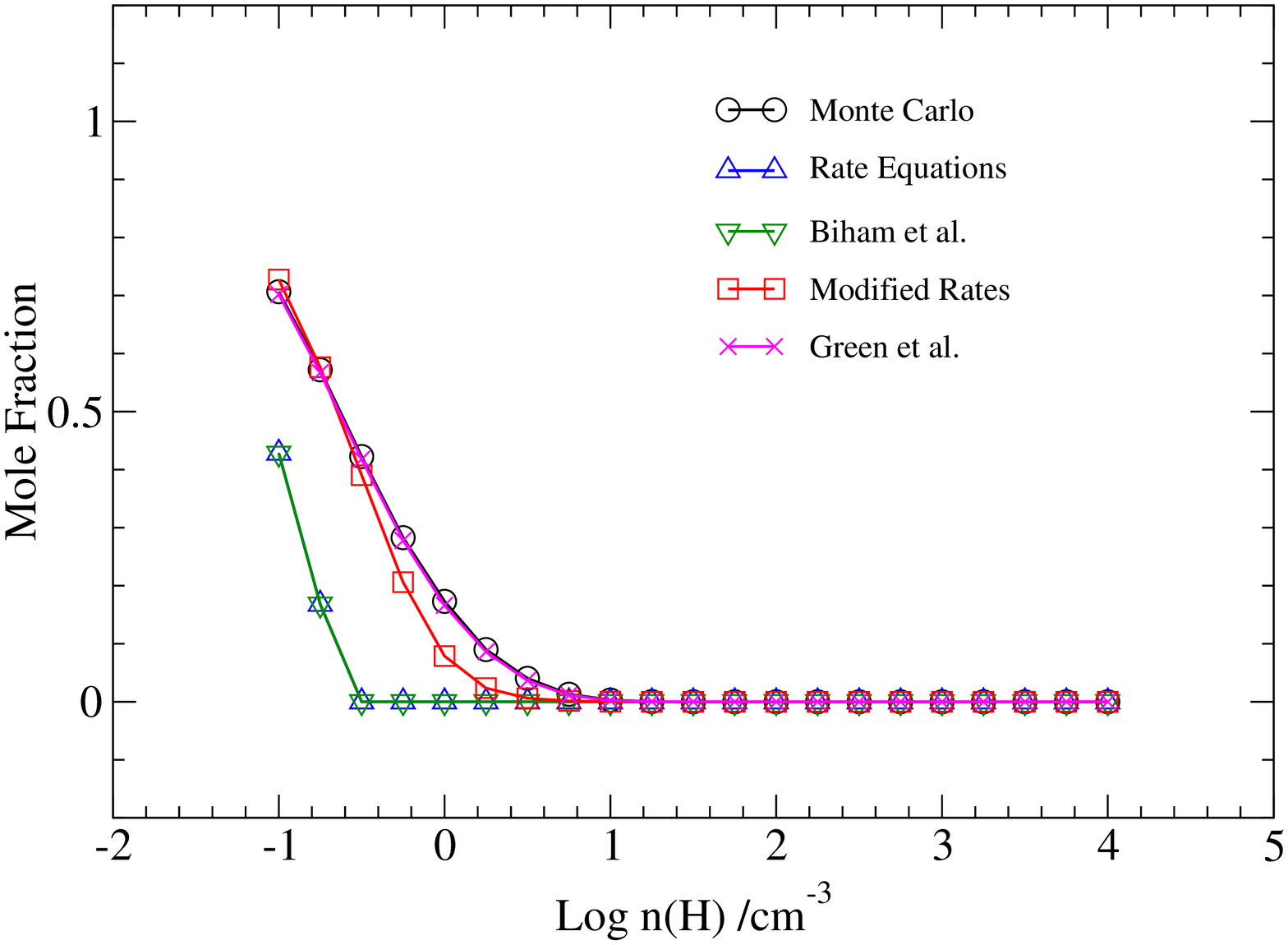}}
          \caption{H,O system. Mole fraction of surface molecular oxygen
          determined via various methods plotted vs the gas-phase concentration
	  of H for a  10 K system.
                  }
             \label{o2}
       \end{figure}

\begin{figure}
       \centering
       \rotatebox{-90}{\includegraphics[width=20 em]{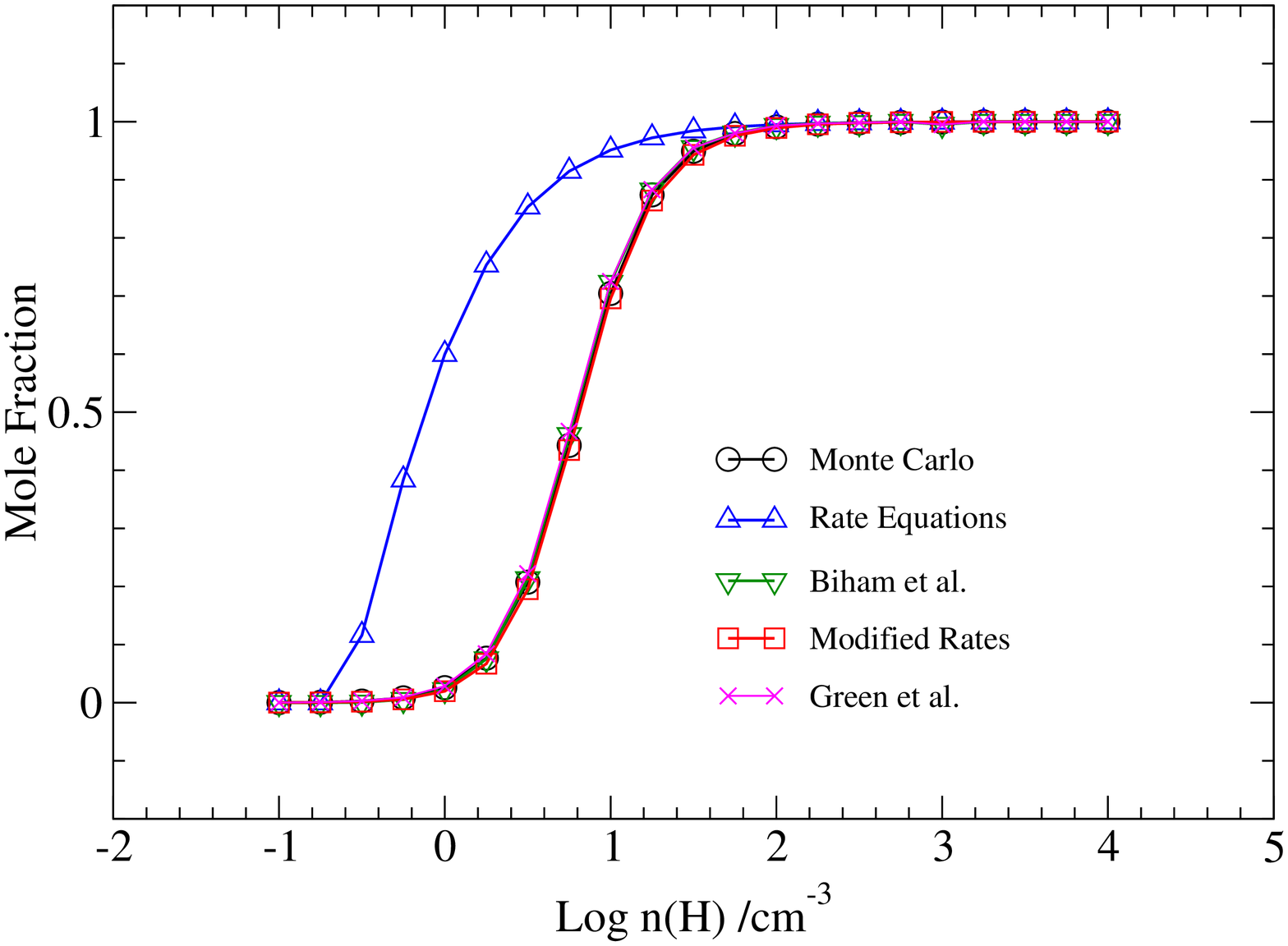}}
          \caption{H,O system. Mole fraction of surface molecular 
hydrogen determined
          via various methods plotted
          vs the gas-phase concentration of H for a 10 K system.
                  }
             \label{h2}
       \end{figure}

\begin{figure}
       \centering
       \rotatebox{-90}{\includegraphics[width=20 em]{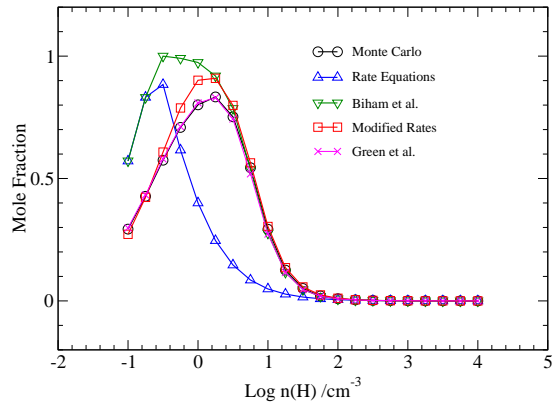}}
          \caption{H,O system. Mole fraction of surface OH determined 
via various methods
	  plotted vs the
          gas-phase concentration of H for a 10 K system.
                  }
             \label{oh}
       \end{figure}

Figs.~\ref{o2}, ~\ref{h2}, and ~\ref{oh} show the mole fractions of
the three diatomic molecules, calculated by the various approaches as
functions of the gas-phase atomic hydrogen abundance.  For the range
of conditions investigated, the exact master equation (Exact ME) and
the Monte Carlo simulation (Appendix) show excellent agreement for all
three species.  The independent probability approximation of Biham et
al.  ~(\cite{bfp2001}) (Approx.  ME) follows the inaccurate results of
the rate equations for O$_{2}$, is in good agreement with the exact
approaches for H$_{2}$, and is in tolerable agreement for OH except at
very low gas-phase H concentrations, where it approaches the
inaccurate rate equation results.  In general, the semi-empirical
modified rate equation approach outperforms the independent
probability approximation.  Biham (private communication) has reported
better success with the approximation for slower diffusion rates, so
one should not rule it out for all situations.

%Another test of whether or not an approximation based on independent
%probabilities is useful is to compare the two-body correlations with
%the products of the average surface abundances, both of which are
%computed with the master equation approach.  In Table \ref{hocorrel},
%we list the average surface abundances and correlations for atomic H
%and O when $n({\rm H})$ = 1 cm$^{-3}$.  It can be seen that
%the correlations are not remotely equal to the products of the
%average concentrations.  For example, the product of the H and O
%abundances is $1.0 \times 10^{-3}$ while the correlation
%$\langle N_{\rm H}N_{\rm O} \rangle$ is $4.8
%\times 10^{-11}$.

Another test of whether or not an approximation based on independent
probabilities is useful is to compare various correlations and
averages computed with both the exact and the approximation master
equation approaches. In Table \ref{hocorrel}, we list the average
surface abundances and correlations for the atomic H and O 
when $n({\rm H})$ = $n({\rm O})$ = 1 cm$^{-3}$. We can
see that the values calculated with both approaches are in mixed
agreement, just as Figs. 1-3 suggest. For example, while the hydrogen
surface abundance is almost identical in both cases, the oxygen
abundances differ from each other by almost seven orders of
magnitude.  Moreover, the correlation $\langle N_{\mathrm{H}}
 N_{\mathrm{O}}  \rangle $ is nowhere near the product of the 
individual averages when calculated by the exact approach.

\begin{table}
\caption{Abundances and correlations for the H, O system}
\label{hocorrel}
\begin{flushleft}
\begin{tabular}{lll}
\hline \hline
      \noalign{\smallskip}
                 Average & \multicolumn{1}{c}{Exact} & 
	\multicolumn{1}{c}{Approx.} \\
                 \hline
                 \noalign{\smallskip}
$\langle \mathrm{N}_{\mathrm{H}} \rangle$  & 6.24(-03) & 5.69(-03) \\
$\langle \mathrm{N}_{\mathrm{O}} \rangle$  & 1.82(-01) & 1.23(-08) \\
$\langle \mathrm{N}_{\mathrm{H}} \rangle 
	\langle \mathrm{N}_{\mathrm{O}} \rangle$ & 
	1.14(-03)  & 7.00(-11)\\
$\langle \mathrm{N}_{\mathrm{H}}\mathrm{N}_{\mathrm{O}} \rangle$ &
	4.80(-11)  & \multicolumn{1}{c}{---} \\
%$\langle \mathrm{N}_{\mathrm{H}}(\mathrm{N}_{\mathrm{H}}-1) \rangle$ &
%	7.03(-13)  & \multicolumn{1}{c}{?} \\
%$\langle \mathrm{N}_{\mathrm{O}}(\mathrm{N}_{\mathrm{O}}-1) \rangle$ &
%	6.75(-03)  & \multicolumn{1}{c}{?} \\
\noalign{\smallskip}
                 \hline \hline
              \end{tabular}
	\end{flushleft}
        \end{table}

%************************
\subsection{H,O,CO system}

For this system, we performed calculations at the three densities
shown in Table \ref{density} using the simple rate equation, Monte
Carlo, and master equation methods.  For the master equation
calculations, the set of minimum $\mathcal{N}_{i}$'s which must be
used is \{2,2,1,1,1\} , since it is necessary to consider at least two
atoms of H and O on the surface for the production of H$_{2}$ and
O$_{2}$, respectively.  Calculations were first performed with this
minimal cutoff for the five minor species, and the results checked by
comparison with the Monte Carlo method, and by increasing the
$\mathcal{N}_{i}$'s.  In general, one can get a picture of what cutoff
is needed for each species by looking at the average number of each
minor species calculated at a given cutoff, or even the number
obtained via the rate equation method.  If this number approaches
unity, then a higher cut-off is needed, and if it exceeds unity, then
it is reasonable to treat the species as a major one not requiring
inclusion in the many-body probability $P$.

Because the
steady-state criterion leads to the fact that
     the surface abundances of O and OH in all three
density cases are equal, we raised the cut-off of OH from 1 to 2
leading to $\mathcal{N}_{i}$'s of
\{2,2,2,1,1\}, which should produce more accurate results.
In the case of high density, however, the O and OH abundances are
sufficiently high (see the detailed discussion below) that we also
tried the case \{2,3,3,1,1\}.  Any further
increase of $\mathcal{N}_{i}$'s led to an increase in the computing
time without changing the results significantly.   After the
following discussion, we
turn to analogous calculations with slow diffusion rates, where the
cutoff problem is more severe.

       \begin{table*}
          \caption{Calculated  populations for surface species at low density
          and 10$^{3}$ yr}
             \label{loden.results}
             \begin{tabular}{lrrrr}
                \hline \hline
                \noalign{\smallskip}
                Species  & Rate eq. & Monte Carlo & Master Eq. & Master
                Eq. \\
			         &          &             & 22111 & 22211 \\
                \hline
                \noalign{\smallskip}
Total      & 1.69(+04) & 1.65(+04) & 1.63(+04) & 1.65(+04) \\
Total (monolayers) &  1.68(-2) & 1.65(-02) & 1.63(-02) & 1.65(-02)\\
H          & 1.21(-05) & 1.00(+00) & 7.96(-03) & 7.96(-03) \\
O          & 5.21(-07) & 0.00(+00) & 1.92(-02) & 1.90(-02) \\
OH         & 5.21(-07) & 0.00(+00) & 1.86(-02) & 1.93(-02) \\
H$_{2}$    & 1.11(+02) & 2.00(+00) & 1.94(+00) & 1.94(+00) \\
O$_{2}$    & 3.64(-07) & 1.70(+02) & 1.62(+02) & 1.62(+02) \\
H$_{2}$O   & 1.03(+04) & 9.90(+03) & 9.65(+03) & 9.86(+03) \\
CO         & 1.51(+02) & 0.00(+00) & 2.81(-02) & 2.81(-02) \\
HCO        & 3.28(-07) & 0.00(+00) & 1.22(-02) & 1.22(-02) \\
H$_{2}$CO  & 1.55(+02) & 0.00(+00) & 2.82(-02) & 2.83(-02) \\
H$_{3}$CO  & 3.28(-07) & 0.00(+00) & 1.23(-02) & 1.23(-02) \\
CH$_{3}$OH & 6.17(+03) & 6.40(+03) & 6.28(+03) & 6.39(+03) \\
CO$_{2}$   & 2.24(-07) & 9.00(+01) & 8.95(+01) & 8.95(+01) \\
{\it CPU} (s) & 0.5 & 11 & 1 & 3 \\
\noalign{\smallskip}
                \hline \hline
             \end{tabular}
%\begin{list}{}{}
%\item[$^{\mathrm{a}}$] Quantum tunelling included
%\end{list}
       \end{table*}

     %  Low densities: normal rates gets major species right.  Lowest cut
      % in good agreement with Monte Carlo, slightly higher cut better.
     %  Charnley right that CO and H2CO agree at steady state. 25\% of a
     %  monolayer

      \begin{table*}
          \caption{Calculated populations for surface species at
intermediate density
          and 10$^{3}$ yr}
             \label{intden.results}
             \begin{tabular}{lrrrr}
                \hline \hline
                \noalign{\smallskip}
                Species  & Rate eq. & Monte Carlo & Master Eq. & Master
                Eq. \\
			         &          &             & 22111 & 22211 \\
                \hline
                \noalign{\smallskip}
Total      & 1.50(+05) & 1.34(+05) & 1.25(+05) & 1.33(+05) \\
Total (monolayers) & 1.50(-1) & 1.34(-01) & 1.25(-01) & 1.33(-01)\\
H          & 5.41(-06) & 1.00(+00) & 3.01(-03) & 2.88(-03) \\
O          & 9.75(-06) & 0.00(+00) & 1.35(-01) & 1.36(-01) \\
OH         & 9.75(-06) & 0.00(+00) & 1.11(-01) & 1.35(-01) \\
H$_{2}$    & 2.20(+01) & 5.00(-01) & 7.32(-01) & 7.01(-01) \\
O$_{2}$    & 1.12(-04) & 9.40(+03) & 8.92(+03) & 9.03(+03) \\
H$_{2}$O   & 8.57(+04) & 6.02(+04) & 5.18(+04) & 5.93(+04) \\
CO         & 3.39(+03) & 1.00(+00) & 7.43(-01) & 7.76(-01) \\
HCO        & 7.36(-06) & 0.00(+00) & 1.14(-01) & 1.14(-01) \\
H$_{2}$CO  & 3.47(+03) & 1.00(+00) & 6.18(-01) & 7.11(-01) \\
H$_{3}$CO  & 7.36(-06) & 0.00(+00) & 1.21(-01) & 1.22(-01) \\
CH$_{3}$OH & 5.79(+04) & 5.79(+04) & 5.79(+04) & 5.81(+04) \\
CO$_{2}$   & 8.23(-05) & 6.60(+03) & 6.64(+03) & 6.64(+03) \\
{\it CPU} (s) & 1 & 13 & 1 & 3 \\
\noalign{\smallskip}
                \hline \hline
             \end{tabular}
%\begin{list}{}{}
%\item[$^{\mathrm{a}}$] Quantum tunelling included
%\end{list}
       \end{table*}

      \begin{table*}
          \caption{Calculated populations for surface species at high density
          and 10$^{3}$ yr}
             \label{hiden.results}
             \begin{tabular}{lrrrrr}
                \hline \hline
                \noalign{\smallskip}
                Species  & Rate eq. & Monte Carlo & Master Eq. & Master
                Eq. & Master eq. \\
			         &          &             & 22111 &
22211 & 23311 \\
                \hline
                \noalign{\smallskip}
Total      & 1.17(+06) & 1.11(+06) & 9.20(+05) & 9.63(+05) & 1.09(+06) \\
Total (monolayers) & 1.17(+00) & 1.11(+00) & 9.20(-01) & 9.63(-01) &
1.09(+00)\\
H          & 3.42(-10) & 0.00(+00) & 1.30(-08) & 1.09(-08) & 8.29(-09) \\
O          & 4.52(-01) & 1.00(+00) & 4.90(-01) & 5.06(-01) & 5.76(-01) \\
OH         & 4.52(-01) & 1.00(+00) & 2.60(-01) & 4.05(-01) & 5.97(-01) \\
H$_{2}$    & 8.79(-8) & 0.00(+00) & 2.90(-06) & 2.50(-06) & 1.89(-06) \\
O$_{2}$    & 2.74(+05) & 2.81(+05) & 1.80(+05) & 1.91(+05) & 2.68(+05) \\
H$_{2}$O   & 2.51(+05) & 1.79(+05) & 1.00(+05) & 1.33(+05) & 1.71(+05) \\
CO         & 6.47(+05) & 5.28(+05) & 4.80(+05) & 4.93(+05) & 5.23(+05) \\
HCO        & 6.71(-04) & 0.00(+00) & 2.10(-01) & 1.88(-01) & 1.53(-01) \\
H$_{2}$CO  & 1.86(+02) & 5.01(+04) & 6.40(+04) & 5.98(+04) & 5.12(+04) \\
H$_{3}$CO  & 3.96(-07) & 0.00(+00) & 5.30(-02) & 4.54(-02) & 3.62(-02) \\
CH$_{3}$OH & 7.33(-02) & 1.10(+04) & 1.90(+04) & 1.56(+04) & 1.17(+04) \\
CO$_{2}$   & 2.04(+02) & 5.82(+04) & 7.90(+04) & 7.15(+04) & 6.01(+04) \\
{\it CPU} (s) & 1 & 30 & 1 & 3 & 10 \\
\noalign{\smallskip}
                \hline \hline
             \end{tabular}
%\begin{list}{}{}
%\item[$^{\mathrm{a}}$] Quantum tunelling included
%\end{list}
       \end{table*}

The calculated surface populations at 10 K for low, medium, and high density
are shown,
respectively, in Tables \ref{loden.results}, \ref{intden.results}, and
\ref{hiden.results} for a time of 10$^{3}$ yr.  In addition
to the individual populations, the total number of surface species is
shown, as is the CPU time utilized for the calculation.
It can be seen that at most one monolayer is built up during the time
of the calculation.  With the normal assumptions, one monolayer of
material corresponds to a fractional abundance with respect to the
total gas density of 10$^{-6}$.  It is to be remembered that only
major species can be detected on grain surfaces (or in subsequent
evaporation into the gas during star formation) so that inaccuracies
in minor species need not pose a critical problem.

With the large diffusion rates, the minor species (H, O, OH, HCO, and
CH$_{3}$O) all have exceedingly low abundances ($<1$ per grain), a
situation known as the `'accretion limit.''  Such
abundances are not capable of being determined accurately by the Monte Carlo
method, which yields integers only.   Occasionally we
have run the simulations several times and averaged the results, so
that non-integers can be obtained. In the
   accretion limit, the
rate equations should not reliably yield accurate answers for the
stable species, and it is easy to see from the tables that such can be
the case here.  Concerning the more accurate
methods, the minimal master equation method (designated 22111; note
that the commas have been removed) takes
about as little CPU time as the rate equation approach and never yields
results more than a factor of two different from the Monte Carlo
approach for species which can be compared.  The latter method is
significantly more computer intensive.  The master equation
calculation with cutoffs 22211 is typically even better, but at
the expense of a factor of three in computer time.  For the high
density case, the oxygen atom abundance is 0.5, and increasing the
cutoffs to 23311 improves the agreement with the Monte Carlo method
but, again, at the expense of increased computer time.  We conclude
that under these physical conditions, the minimal master equation
method is a fast and reliable appoach for the H, O, CO system.
Although the results for the modified rate method were not presented,
this approach typically does better than the simple rate method but
worse than the other approaches.

The tabulated results are all for a temperature of 10 K. We have also
done calculations for high density at temperatures through 20 K. The
results are plotted for major species in Figs.~ \ref{xmole1} and
\ref{xmole2} in terms of mole fractions vs temperature. Cutoffs of
23311 have been used.  Although
the major species have surface abundances that increase linearly with
time, the mole fractions remain constant.   Results are shown
for the Monte Carlo (MC), master equation (ME), simple rate equation
(RE), and modified rate equation (MR) approaches.  An analogous plot
based on the Monte Carlo method is shown in Caselli et
al.~({\cite{css2002}) and is in a good agreement with the present
results.  The Monte Carlo and master equation results are essentially
identical, while the rate equation results are at best in mixed
agreement with the two exact approaches.  The modified rate method is
significantly better than the simple rate approach but still can show
factor of 2 or greater disagreement with the Monte Carlo and master
equation methods.

Generally speaking, the results show that under the high density
conditions considered here, the production of methanol is reasonably
efficient only at the lower temperatures considered.
Its efficiency at 10 K appears to peak at intermediate densities (see
Table \ref{intden.results}).  The mole fraction of CO$_{2}$ also
decreases strongly at the higher temperatures, while O$_{2}$ and CO
tend generally to increase with increasing temperatures.

      \begin{figure}
       \centering
       \rotatebox{-90}{\includegraphics[width=35 em]{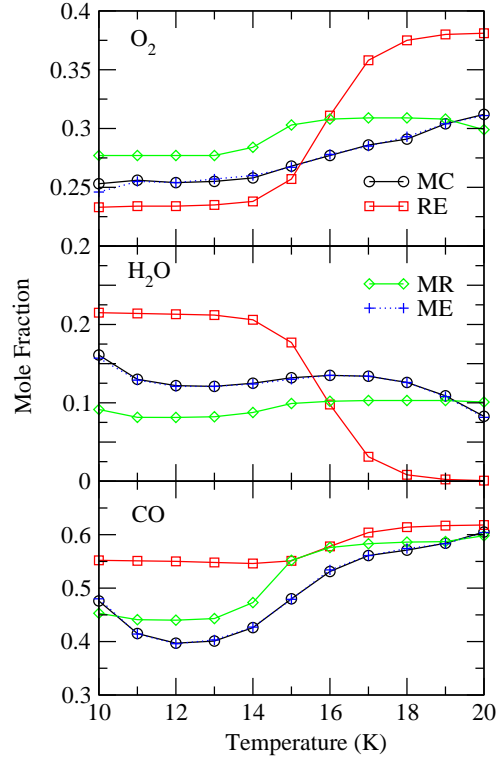}}
          \caption{The mole fractions of surface O$_{2}$, H$_{2}$O, and CO
          after 1000 yr for
          high density conditions plotted vs. temperature (K)}
             \label{xmole1}
       \end{figure}

      \begin{figure}
       \centering
       \rotatebox{-90}{\includegraphics[width=35 em]{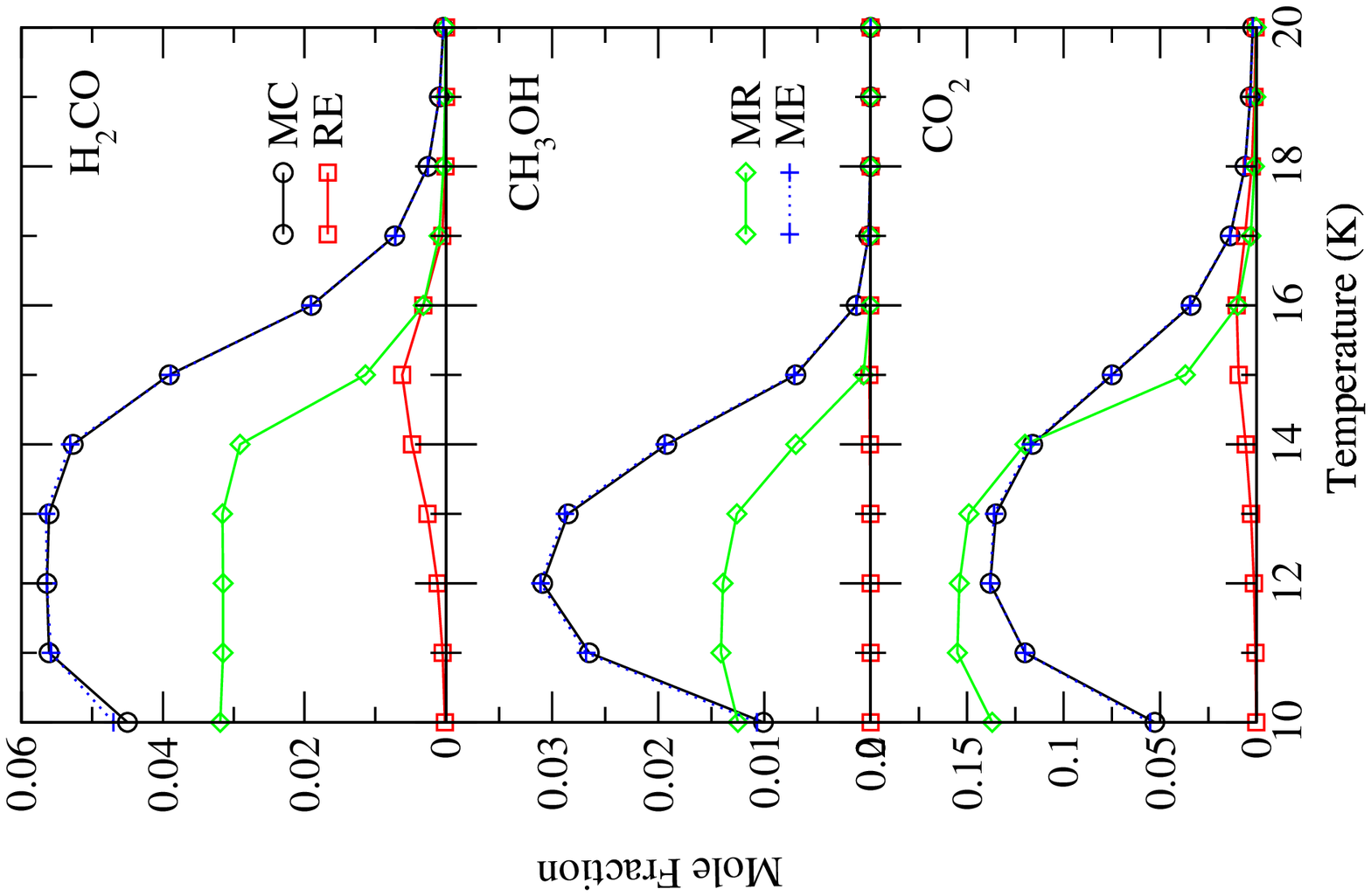}}
          \caption{The mole fractions of surface H$_{2}$CO, CH$_{3}$OH, and
          CO$_{2}$ after 1000 yr for high density conditions plotted vs.
temperature (K)
                  }
             \label{xmole2}
       \end{figure}

\subsection{Slow Diffusion Rates for the H,O,CO system}

As the diffusion rates are lowered, the abundances of surface oxygen
atoms and OH radicals increase dramatically and the need for a more detailed
treatment than the rate equation approach lessens.  Indeed, as the
average abundance of a minor species becomes greater than unity, it
becomes difficult for the many-body master equation treatment to
converge, so that it is best to remove those species from the
many-body probability Eq.~(\ref{prob}) and compute them via
equations similar to Eq.~(\ref{reH2}). In Tables
\ref{lodenslow} and \ref{hidenslow} we plot results at low and high
densities obtained with
the ``slow'' (M2) diffusion rates of Ruffle \& Herbst (\cite{rh2000})
using the rate equation, Monte Carlo, and master equation methods.
For low density, we report results for both the minimal cutoff case
and for a calculation, designated 2xx11 in which O and OH are not treated
probabilistically.  For high density, we can only perform the latter
calculation since the surface abundances for O and OH are very
large.

           \begin{table*}
           \caption{Calculated populations for surface species with ``slow''
           diffusion rates at low density
           and 10$^{3}$ yr}
              \label{lodenslow}
              \begin{tabular}{lrrrr}
                 \hline \hline
                 \noalign{\smallskip}
                 Species  & Rate eq. & Monte Carlo & Master Eq. & Master
                 Eq. \\
			         &          &             & 22111 & 2xx11 \\
                 \hline
                 \noalign{\smallskip}
Total      & 1.68(+04) & 1.69(+04) & 6.98(+03) & 1.68(+04) \\
Total (monolayers) & 1.68(-02) & 1.69(-02) & 6.98(-03) & 1.68(-02)\\
H          & 8.22(-02) & 0.00(+00) & 8.43(-02) & 8.15(-02) \\
O          & 3.75(+00) & 3.00(+00) & 7.86(-01) & 3.79(+00) \\
OH         & 3.75(+00) & 5.00(+00) & 1.81(-01) & 3.79(+00) \\
H$_{2}$    & 1.71(-07) & 0.00(+00) & 1.60(-07) & 1.50(-07) \\
O$_{2}$    & 8.84(-10) & 0.00(+00) & 2.66(-11) & 9.00(-10) \\
H$_{2}$O   & 1.03(+04) & 1.03(+04) & 5.05(+02) & 1.03(+04) \\
CO         & 6.47(+03) & 6.59(+03) & 6.47(+03) & 6.47(+03) \\
HCO        & 1.15(-04) & 0.00(+00) & 3.93(-05) & 7.38(-05) \\
H$_{2}$CO  & 1.58(-01) & 0.00(+00) & 5.49(-02) & 9.96(-02) \\
H$_{3}$CO  & 2.76(-09) & 0.00(+00) & 3.27(-10) & 1.11(-09) \\
CH$_{3}$OH & 2.52(-06) & 0.00(+00) & 3.04(-07) & 1.00(-06) \\
CO$_{2}$   & 1.36(-14) & 0.00(+00) & 1.24(-15) & 8.77(-15) \\
{\it CPU} (s) & 0.4 & 12 & 1.2 & 0.2 \\
\noalign{\smallskip}
                 \hline \hline
              \end{tabular}
%\begin{list}{}{}
%\item[$^{\mathrm{a}}$] Quantum tunelling included
%\end{list}
        \end{table*}

For the slow rates, it can be seen that little chemistry occurs except for the
formation of water and, at high densities, OH. (Any H$_{2}$ formed
has desorbed.)  Nevertheless, the simple rate equation approach agrees
surprisingly well with the 2xx11 reduced master equation approach.
This occurs despite the fact that the average H atom abundance is less
than unity, presumably because the abundances of its reactive partners
O and OH exceed unity.
The Monte Carlo method is also in fine agreement for major species
although it once again is more computer intensive; it
cannot really be compared with the other methods for minor species
given the large statistical uncertainties in the small numbers.  For
low density, the minimal cutoff master equation approach is
unreliable, as is to be expected when the O abundance exceeds unity.
For high density, the O atom abundance is so large that a calculation
with astronomical significance should include the Eley-Rideal
mechanism of surface chemistry, in which gas-phase species collide
reactively with nearly stationary species on grain surfaces.

          \begin{table*}
           \caption{Calculated populations for surface species with ``slow''
           diffusion rates at high density
           and 10$^{3}$ yr}
              \label{hidenslow}
              \begin{tabular}{lrrr}
                 \hline \hline
                 \noalign{\smallskip}
                 Species  & Rate eq. & Monte Carlo & Master Eq. \\
			         &          &             & 2xx11  \\
                 \hline
                 \noalign{\smallskip}
Total      & 1.45(+06) & 1.45(+06) & 1.45(+06)  \\
Total (monolayers) & 1.45(+00) & 1.45(+00) & 1.45(+00) \\
H          & 2.33(-05) & 0.00(+00) & 2.33(-05)   \\
O          & 4.50(+05) & 4.51(+05) & 4.50(+05)   \\
OH         & 1.97(+05) & 1.97(+05) & 1.97(+05)   \\
H$_{2}$    & 1.38(-14) & 0.00(+00) & 1.38(-14)  \\
O$_{2}$    & 4.25(+00) & 4.00(+00) & 4.25(+00)   \\
H$_{2}$O   & 1.52(+05) & 1.52(+05) & 1.52(+05)   \\
CO         & 6.47(+05) & 6.48(+05) & 6.47(+05)   \\
HCO        & 5.04(-03) & 0.00(+00) & 5.01(-03)   \\
H$_{2}$CO  & 3.88(-03) & 0.00(+00) & 3.87(-03)   \\
H$_{3}$CO  & 2.94(-11) & 0.00(+00) & 2.93(-11)   \\
CH$_{3}$OH & 2.19(-11) & 0.00(+00) & 2.17(-11)   \\
CO$_{2}$   & 4.75(-08) & 0.00(+00) & 4.73(-08)   \\
{\it CPU} (s) & 0.6 & 27 & 0.6 \\
\noalign{\smallskip}
                 \hline \hline
              \end{tabular}
%\begin{list}{}{}
%\item[$^{\mathrm{a}}$] Quantum tunelling included
%\end{list}
        \end{table*}

\section{Discussion}

We have shown that a moderately complex network of chemical reactions
that occur diffusively on the surfaces of dust particles can be
studied successfully by a master equation approach previously used
only for significantly simpler systems (Biham et al.~\cite{bfp2001};
Green et al.~\cite{gtp2001}).  It is important to use an ``exact''
method such as the master equation approach or its Monte Carlo
realization when there is a very low surface abundance of reactive
species, since the rate equation method may be inaccurate, and the
semi-empirical modified rate equation method may not be entirely
correct either.  The advantage of the master equation approach to the
Monte Carlo method is that the former involves the solution of
simultaneous differential equations.  It is therefore facile to
consider the gas-phase chemistry occurring simultaneously, because
gas-phase abundances are also determined by solving simultaneous
differential equations.  Moreover, it is possible to determine when
the master equation method must be used for all minor species, and
when it is acceptable to use the simpler and faster rate equation
approach to diffusive surface chemistry.  Specifically, when the
abundances of reactive species on grain surfaces begin to exceed
unity, there may be no need to use the more detailed approach, as can
easily be tested.  Calculations reported here show instances where even if
the average atomic hydrogen surface abundance is below unity, the
simple rate equation method works well if the O and OH abundances are high.
Finally, for the system studied here, the master equation approach is
actually faster than its Monte Carlo analog.

With all of these advantages, one might conclude that it should be
facile to implement the master equation approach in current complex
gas-grain models of interstellar clouds.  But this optimism is
misplaced.  If the master equation method discussed in this paper is
to be extended to still larger systems of reactions, such as that used
in current gas-grain chemical models (Ruffle \& Herbst \cite{rh2000}),
some method must be found to reduce the number of simultaneous
equations neccessary.  Let us consider the extent of the problem.  The
minimal-cutoff (22111) master-equation approach for fast diffusion
rates requires the simultaneous solution of $3 \times 3 \times 2
\times 2 \times 2 = 72$ simultaneous differential equations to fully
determine the many-body $P$, excluding the additional coupled
equations for the major species.  As one increases the number of minor
species, the number of simultaneous equations to be solved increases
dramatically.  Suppose, we wish to consider the chemistry of deuterium
fractionation in the H, O, CO network discussed here.  Such an
extension requires the following new minor species: D, OD, DCO,
H$_{2}$DCO, HD$_{2}$CO, D$_{3}$CO, making a total of 11 such species.
The deuterium fractionation in this model was treated successfully by
the Monte Carlo method (Caselli et al.~\cite{css2002}).  Assuming that
minimal cutoffs of 1 are needed for all these additional species with
the exception of D, we calculate that a total number of 6,912
simultaneous equations is needed for the many-body probability
function $P$.  This compares unfavorably with the total of 652
equations used for both the gas-phase and surface chemistry in our
most complex models.  It is difficult to even load the variables for
such a calculation onto most computers, and the computer time
necessary is virtually prohibitive.  Even if only D is treated
stochastically and OD, DCO, etc.  are treated via rate equations, the
computer time increases by a factor larger than 10 compared with the
H,O,CO system.

Given the importance of developing approximation methods, we have
investigated a simple such approximation: the idea that the many-body
probability can be approximated as the product of individual,
independent probabilities.  We have seen that, even for the simple O,H
system, the approximation of independent probabilities is not a
reliable approach to the solution of the many-body master equation if
one assumes fast diffusion rates.  Although we did not report the
results here, our extension of the independent probability approach to
the more complex H,O,CO system has also met with failure.  We are
currently studying other approximation methods.  One rather promising
approach at this time is to limit the total number of equations by
limiting the total number of reactive species on a grain surface.  We
hope to report results with this method in the near future.

\begin{acknowledgements}
     The Astrochemistry Program at The Ohio State
University is supported by The National Science Foundation (US).
V. I. S. acknowledges support from grant RFBR 01-02-16206.  We thank
the Ohio Supercomputer Center for time on their Cray SV1 machine.
\end{acknowledgements}

\appendix

\section{}

\subsection{Kinetics of Chemically Reacting System}

        Let us consider an interstellar gas consisting of $K$ different
        atomic and molecular species labelled by
$\alpha_i, i=1,..., K \ge 1$
in a fixed physical volume $V$ with concentration $n_{\alpha_{i}}$.
Each particle of
     species $\alpha_i$ is characterized by mass
$m_i$, velocity ${\bf c}_i$, and a set ${\bf z}_i$ of internal
quantum states.
The chemical species in the gas can interact through
$m=1,...,M \ge 1$ chemical reactions of the type:
\begin{equation}
\label{eq02}
m: \alpha_i({\bf c}_i,{\bf z}_i) + \alpha_j({\bf c}_j,{\bf z}_j)
\rightarrow \alpha_k({\bf c}_k,{\bf z}_k) +
\alpha_l({\bf c}_l,{\bf z}_l).
\end{equation}
The probability that reaction (\ref{eq02}) occurs at
        a specific relative velocity with products scattered in a certain
        direction is related to
\begin{equation}
\label{eq28}
g_{ij}d\sigma_{m}=g_{ij}\frac{d\sigma_{m}(g_{ij},\Omega)}{d\Omega}d\Omega~~,
\end{equation}
where $d\sigma_{m}$ is the reactive differential scattering cross-section,
$g_{ij}=\vert{\bf c}_i-{\bf c}_j\vert$
is a relative velocity, and $\Omega$ is a solid scattering angle.
The differential cross section depends on the interaction potential
of the particles involved and
can be calculated by methods of quantum mechanics or measured in laboratory
experiments (Light et al.~\cite{lrs1969}).

In astrochemical problems, interstellar
gases are usually considered under the assumption of local
thermal equilibrium.  This requires that reactive collisions occur
less frequently than elastic and inelastic processes so that a
temperature can be maintained.  At thermal equilibrium, the
distribution of possible molecular
speeds is given in three dimensions by the Maxwellian distribution
function $f_{\alpha}^{(M)}$. It is then possible
     to obtain the conventional rate equations of chemical kinetics
     (Light et al. \cite{lrs1969}):
\begin{equation}
\label{eq91a}
\frac{d}{d t} n_{{\alpha}_i}(t)=
\sum\limits_{m}  [n_{{\alpha}_k} n_{{\alpha}_l} q_{{\alpha}_k,{\alpha}_l}
~-~n_{{\alpha}_i} n_{{\alpha}_j}q_{{\alpha}_i,{\alpha}_j}]~~,
\end{equation}
where the functions $q_{..}$ are kinetic rate coefficients for the
forward (with cross section indicated by $\rightarrow$)
and backward (with cross section indicated by $\leftarrow$) directions
of chemical reaction \ref{eq02}:
\begin{equation}
	\begin{array}{l}
q_{{\alpha}_k,{\alpha}_l}=\int \,d{\bf c}_k \,d{\bf c}_l
g_{kl}\,d \sigma_{m\leftarrow} f_{\alpha_k}^{(M)}({\bf c}_k)
f_{\alpha_l}^{(M)}({\bf c}_l) \\
q_{{\alpha}_i,{\alpha}_j}
=\int \,d{\bf c}_i \,d{\bf c}_j
g_{ij}\,d \sigma_{m\rightarrow} f_{\alpha_i}^{(M)}({\bf c}_i)
f_{\alpha_j}^{(M)}({\bf c}_j)~~.
\end{array}
\end{equation}

\subsection{Stochastic Approach to a Chemically Reacting System}

        Chemical kinetics in a rarefied interstellar gas
can be formulated as a stochastic evolution of an ensemble of
atoms and molecules.  A stochastic approach is based
on the relationship between two basic ways of describing the
chemically reacting and evolving gas: (i)
the Liouville dynamic equation and corresponding kinetic equations
(Smith \cite{s1969}) and (ii) the stochastic
laws, describing a random process and its stochastic Kolmogorov
equation equivalent
(Gillespie \cite{gdt1976}; Marov et al. \cite{metal1997}).
For a space-uniform gas, the changes in the state of the gas caused by instant
collisions can be considered to be
jump-like Markovian processes (Marov et al. \cite{metal1997}), after
which the state of the system does not contain the memory of how the
state was reached.

In a stochastic treatment,
the evolution of the reacting system is governed by the so-called
chemical master equation (Gillespie \cite{gdt1982}; van Kampen
\cite{vk1992}):
\begin{equation}
{\partial \phi({\bf N},t)\over\partial t}=
\sum\limits_{m=1}^M [ a_m({\bf N}^m)\phi({\bf N}^m,t)
-a_m({\bf N})\phi({\bf N},t)]~~,
\label{eq05}
\end{equation}
which is linear with respect to the probability density $\phi({\bf N},t)$
that a system is described by state ${\bf N}$ at time $t$.   The
parameters
in the equation are defined below:
\begin{itemize}
\item[] a)  The state of the system is
characterized by
\begin{equation}
{\bf N}(t) = \{N_1(t),...,N_K(t) \}~~,
\label{eq06}
\end{equation}
where the atomic and molecular populations $N_i(t)$ for each species
$i$ are random integer
variables in the considered gas volume $V$ at time $t$;

\item[] b) The state of the system realized after an instant change
of molecular populations
in accordance with the stoichiometric scheme of reaction $m$ is given by
\begin{equation}
\begin{array}{l}
{\bf N}^m = \{..., N_i-1,...,\\
\; \; \; \; N_j-1,...,N_k+1,...,N_l+1,...\};
\label{eq07}
\end{array}
\end{equation}

\item[] c)
The probability that a certain reaction $m$ will take place in an
infinitesimal time interval $[t,t+dt]$ is given by the expression
$a_m({\bf N})dt$,
where $a_m({\bf N})$ is independent of $dt$ and is equal to:
\begin{equation}
\label{eq08a}
a_m({\bf N})=V^{-1}h_{ij}q_{m}.
\end{equation}
Here, $q_{m}$ is the rate coefficient for reaction $m$, and  $h_{ij}$ is a
combinatorial factor equal to the number of possible pairs of
reacting molecules, or $N_iN_j$, for
$\alpha_i \neq \alpha_j$, and  $N_i(N_i-1)/2$ for $\alpha_i = \alpha_j$.

The function $a_m({\bf N})$ depends on
the specific chemical channel $m$, the current gas state ${\bf
N}={\bf N}(t)$, and
the gas temperature and volume. This function, usually called
the propensity function (Gillespie \cite{gdt1976}), refers to
processes that lead away from state ${\bf N}$.  The function
$ a_m({\bf N}^m)$ depends on the state of the gas ${\bf N}^{m}$, and
refers to processes that lead to state ${\bf N}$.
\end{itemize}

     Because ${\bf N}(t)$ refers to a Markovian jump-like random
process,
the time distribution between collisions is given by an exponential law:
\begin{equation}
\label{eq09}
P \{ \tau({\bf N} \rightarrow {\bf N}^\prime) \leq \tau \} = 1-
\exp{(-a_0({\bf N}) \tau)}~~,
\end{equation}
where $a_{0}$, the total reaction probability, is defined by
\begin{equation}
a_0({\bf N})=\sum\limits_{m=1}^Ma_m({\bf N}).
\label{eq10}
\end{equation}

Expressions (\ref{eq06}) - (\ref{eq10}) give an exact definition of the
random state of the system ${\bf N}(t)$ describing a chemically reacting
gas within a stochastic framework. How do we describe a system of
reactions on the surface of a dust particle?

It is well known that the random nature of interstellar grain
surface chemistry, as well as the accretion and desorption processes leading
to  grain mantle
growth, necessarily involves a stochastic framework (Tielens \& Hagen
\cite{th1982}; Tielens
\& Charnley \cite{tc1997};
Herbst \cite{eh2000a}; Charnley \cite{c2001}).
The stochastic treatment of grain surface chemistry can be formulated in
terms of the master equation approach. To do this, the
surface should be represented by a lattice or a monolayer - each lattice
point corresponds to a surface site.  A lattice point can assume a number
of distinct values that stand for various adsorbed molecules (with zero
for a free
site). The monolayer, together with all its site populations, is considered
as a state vector ${\bf N}$ for the surface reacting system. Each surface
reaction changes the monolayer population vector in accordance with
the reaction stoichiometry.  The evolution of
this surface reacting system over time is described by a chemical
master equation - Eq.~\ref{eq05}).  Since there are a large
number of reactive sites on interstellar dust particles, additional
assumptions are normally made for stochastic theories - that all sites
are identical, that one need only follow the number of particles of
a given species on the entire grain, and that the distribution of
particles on a grain is random.

In astrochemical environments at very low
temperatures ($\sim $ 10~K), light atoms are the major mobile species on
grain surfaces.
Atoms migrate mainly by thermal hopping from site to site with a timescale
$$
      \tau_{\rm h}^{\alpha}=\nu^{-1}\exp(E_{\rm D}^{\alpha}/kT_{\rm d}),
$$
where $E_{\rm D}^{\alpha}$ is the energy barrier for surface diffusion for
atomic species ${\alpha}$, $T_{\rm d}$ is the
surface temperature, and $\nu$ is the vibrational frequency of the particle
in the lattice binding site ($\sim~10^{12}$ s$^{-1}$). Light hydrogen
atoms  can also migrate by quantum mechanical tunneling, with a
characteristic timescale $\tau_{\rm H}~\sim~10^{-12}$ s.
Migration on the surface leads to reactions with other
light migrating atoms or heavy and relatively static molecules and radicals
with reactive transition probabilities
\begin{equation}
\label{eq17a}
a_m({\bf N})=h^{\prime}_{ij}\times p_m \times (\tau^{-1}_{\alpha} +
\tau^{-1}_{\beta}),
\end{equation}
where the $\tau^{-1}$ factors are for the two reacting species and can
be either for surface hopping or tunneling.  The factor $p_m$ is unity
unless there is an activation energy barrier
$E_m$.   This factor is then
equal to the Arrhenius factor $\exp(-E_m/kT_{\rm d})$ or, for the case of
quantum mechanical tunneling through the potential barrier of height
$E_m$ and width
$L_m$, is equal to $\exp(-4\pi L_m (2\mu E_m)^{0.5}/h)$, where $\mu$
is the reduced mass.  The prime in
the combinatorial factor
$h^{\prime}_{ij}$ means that there is a denominator with the actual
number of sites
on a grain.  The denominator converts the rate of diffusion from one
site to another into the rate of diffusion $t^{-1}_{\rm diff}$ over
the equivalent of an entire grain.
As suggested by Charnley (\cite{c2001}), the surface chemical network should be
extended by interpreting the accretion and desorption processes as additional
reactions responsible for the chemical coupling between gas-phase and
grain mantle
fractions of the interstellar gas.

\subsection{Methods for Solving the Chemical Master Equation}

\subsubsection{ Monte Carlo Algorithms}

These algorithms are based on the fact that the
probability of generating the stochastic ``trajectory'' with a Monte Carlo
algorithm is exactly the probability that would come out of the solution
of the corresponding master equation.

To accomplish this, the homogeneous Markovian process ${\bf N}(t)$
is replaced by an equivalent uniform Markovian chain.
An exact realization for the Markovian chain on a discrete time grid is as
follows. We choose a
time interval $\triangle t$ and determine times
$t_1 = 1\triangle t$, $t_2 = 2 \triangle t$, \ldots, $t_{\delta} =
\delta \triangle
t$, for
which we will store the values of {\bf N}$_{1}$ = {\bf N}($t_{1}$),
{\bf N}$_{2}$ = {\bf N}($t_{2}$),\ldots, {\bf N}$_{\delta}$ = {\bf
N}($t_{\delta}$),
respectively.
%Let us say that there have already occurred $s-1$ transitions and the
%time elapsed for these transitions has been determined to be
%$T^{(s-1)}$. Also, let $\delta$ be so that $t_{\delta} \leq T^{(s-1)} <
%t_{\delta+1}$; that is, we have already determined the values of {\bf
%N}$_{1}$, {\bf N}$_{2}$, \ldots, {\bf N}$_{\delta}$.
To determine the state {\bf N}$_{\delta+1}$ of a dust particle,
we perform the following steps:

\begin{itemize}
\item[] a) we determine randomly: (1) which reaction will occur during the
transition via (\ref{eq10}) and (\ref{eq17a}) , and (2) in accordance
with the probability distribution
(\ref{eq09}), the time $\tau$ that has elapsed from the previous
transition. This can be done using the {\it direct simulation} or
{\it first reaction} methods (Gillespie \cite{gdt1976};
Lukken et al.~\cite{lea1998}; Charnley \cite{ch1998}, \cite{c2001}),
the former of which is used here.
Then, we update the species' populations, and advance the
transition time counter as $T^{(s)}=T^{(s-1)}+{\tau}$~;
\item[] b) if the transition time counter satisfies the
following condition:
\[ t_{\delta+1} \leq T^{(s)} <t_{\delta+2}, \]
it means that we have stepped over the next point on the time grid.
In such a case, we assign a value to {\bf N}$_{\delta+1}$ that is equal to the
current state. If the inequality does not hold, we repeat the
operations from step a).
\end{itemize}

We repeat the whole procedure until
the time counter reaches a fixed time T$_{final}$.

Algorithmic steps (a) - (b) represent the exact Monte-Carlo procedure
for solving the chemical master equation.  One realization for the
Markovian chain {\bf N}$_{1}$, {\bf N}$_{2}$, {\bf N}$_{3}$, \ldots,
gives only one possible evolution of the system.  Thus, it is
important that the procedure be repeated and many trajectory
realizations be obtained.  Due to the linearity of the chemical master
Eq.~(\ref{eq05}) and, consequently, of analogous Monte-Carlo
algorithms for its solution, the evolution can be calculated by
averaging through trajectory realizations of the random process ${\bf
N}(t)$.

\subsubsection{Direct Solution}

For surface chemistry, one natural way to deal with the stochastic
approach is to create one
probability variable for each possible state of the reacting
chemical system under study.
If the system can be contained within a limited set of
possible states so that the number of molecules of each species
is limited by some fixed value $\bar{N}_i$ during the system evalution -
$N_i(t) \le \bar{N}_i$ - it is useful to adopt as a representation of
the state probability distribution $\phi({\bf N},t)$ a set of
many-body probabilities $P$ for specific numbers of molecules for the
species being considered.
By substituting this set into the chemical master equation,
we can obtain a set of coupled differential
equations for the time derivative of the detailed probabilities of all
possible states - ${{\rm d}\over{\rm d}t}P(i_1,...i_N)$.
%\begin{equation}
%\left\{ \begin{array}{l}
%{{\rm d}\over{\rm d}t}p(i_1,...i_N)= \\ \\
%\sum\limits_m^M c_m[
%(i+1)(j+1)p(...,i+1,...j+1,...,k-1,...,l-1,...) - \\ \\
%{(i)(j)p(...,i,...j,...,k,...,l,...)]~~.}
%\end{array} \right.
%\label{eq59}
%\end{equation}

Such a set of differential equations is used in this paper, but with
a caveat.
For surface chemistry, it is normally not necessary to include all
species in the model in the realization of the master equation.
Generally, only a few reactive species have surface abundances so low
that a stochastic treament is necessary.
All other molecules  can be described
by a deterministic approach with coventional rate equations.
These two (stochastic and deterministic) subsets of the system are coupled
through the combinatorial factors
\begin{equation}
\bar{h}_{ij}= \langle N_{j} \rangle
\sum\limits_{i=1}^{{\bar{N}_i}}iP(...,i,...)
\end{equation}
where $\langle N_{j} \rangle$ is a mean molecular population for the
deterministic
$j$ species and the $P(...,i,...)$ are probabilities for the stochastic $i$
species.
This division allows us to combine the direct solution of the chemical
master equation for stochastic species with simple or modified rate equations
for deterministic
species to produce realistic grain-surface chemical networks.

\end{document}